\documentclass{article}

\usepackage[english]{babel}

\usepackage[letterpaper,top=2cm,bottom=2cm,left=3cm,right=3cm,marginparwidth=1.5cm]{geometry}
\usepackage{subcaption}
\usepackage{graphicx,booktabs,tabulary,tabularx}
\usepackage{multirow} 
\usepackage{amsmath}
\usepackage{graphicx}
\usepackage[colorlinks=true, allcolors=blue]{hyperref}
\usepackage{makecell}
\usepackage{comment}
\title{A Novel Multimodal System to Predict Agitation in People with Dementia Within Clinical Settings: A Proof of Concept}
{\author{Abeer Badawi $^1$, Somayya Elmoghazy$^1$, Samira Choudhury$^2$ $^3$, Sara Elgazzar $^4$,\\ Khalid Elgazzar$^1$, and Amer Burhan $^2$ $^3$\\
			$^1$ IoT Research Laboratory, Ontario Tech University, Oshawa, ON, Canada\\
            $^2$ Ontario Shores Centre for Mental Health Sciences, Whitby, ON, Canada \\
            $^3$ Temerty Faculty of Medicine, University of Toronto, Toronto, ON, Canada \\
            $^4$ Faculty of Science, Ontario Tech University, Oshawa, ON, Canada}}

\begin{document}
\maketitle

\begin{abstract}
Dementia is a neurodegenerative condition that combines several diseases and impacts millions around the world and those around them. Although cognitive impairment is profoundly disabling, it is the noncognitive features of dementia, referred to as Neuropsychiatric Symptoms (NPS), that are most closely associated with a diminished quality of life. Agitation and aggression (AA) in people living with dementia (PwD) contribute to distress and increased healthcare demands. Current assessment methods rely on caregiver intervention and reporting of incidents, introducing subjectivity and bias. Artificial Intelligence (AI) and predictive algorithms offer a potential solution for detecting AA episodes in PwD when utilized in real-time. We present a 5-year study system that integrates a multimodal approach, utilizing the EmbracePlus wristband and a video detection system to predict AA in severe dementia patients. We conducted a pilot study with three participants at the Ontario Shores Mental Health Institute to validate the functionality of the system. The system collects and processes raw and digital biomarkers from the EmbracePlus wristband to accurately predict AA. The system also detected pre-agitation patterns at least six minutes before the AA event, which was not previously discovered from the EmbracePlus wristband. Furthermore, the privacy-preserving video system uses a masking tool to hide the features of the people in frames and employs a deep learning model for AA detection. The video system also helps identify the actual start and end time of the agitation events for labeling.  The promising results of the preliminary data analysis underscore the ability of the system to predict AA events. The ability of the proposed system to run autonomously in real-time and identify AA and pre-agitation symptoms without external assistance represents a significant milestone in this research field.
\end{abstract}


\section{Introduction}
Dementia is a neurodegenerative condition that leads to a progressive decline in cognition and is one of the leading causes of death, disability, and hospitalization in Canada and worldwide. Currently, dementia is the seventh cause of death worldwide~\cite{WHO2023}. Globally, over 55 million individuals are living with dementia; as the ratio of older people increases, this number will grow to 78 million by 2030 and 139 million by 2050, making dementia a major global health crisis~\cite{WHO2023}. In addition to cognitive and functional decline, people living with dementia (PwD) also experience non-cognitive neuropsychiatric symptoms (NPS) during their illness~\cite{Lyketsos2002}. NPS commonly includes agitation, aggression, apathy, symptoms of psychosis, delusions, hallucinations, and disturbances of sleep and appetite. Among NPS, agitation and aggression (AA) occur frequently in severe cases and are a common source of distress for patients and caregivers~\cite{Ballard2013}. They commonly occur during care and are believed to be manifestations of perceived or real unmet needs~\cite{Ballard2013}. Behaviors of AA include pacing, rocking, gesturing, restlessness, shouting, scratching, throwing objects, and destroying property~\cite{seitz2010prevalence}. These symptoms are the leading cause of hospitalizations, extended length of stay as inpatients, and increased demand for placement in long-term care facilities~\cite{knuff2019use}. AA enormously burdens PwD, their families, caregivers, and healthcare systems. 

In current practices, AA are commonly assessed through caregiver reports. Many observational methods have been developed, including the Neuropsychiatric Inventory (NPI)~\cite{cummings1994neuropsychiatric} and the Cohen-Mansfield Agitation Inventory (CMAI)~\cite{cohen2012cohen}. These assessments are based on manual observations, which are subject to potential bias depending on the caregiver's memory or emotional state. It is possible to address these limitations by using Artificial Intelligence (AI) and predictive algorithms to predict episodes of AA in PwD before they occur. By 2025, AI technologies are expected to be worth an estimated  \$36 billion (US)~\cite{Markets2022}. There is growing evidence that combining AI and sensory technologies to develop a solution for NPS detection will guide the provision of personalized interventions for PwD~\cite{bharucha2009intelligent, Recognition2022, khan2018detecting, fabrizio2021artificial}. The timely detection of critical events in PwD using digital technologies is gaining wide acceptance. For example,  smartwatches are being used to help people with dementia ~\cite{goerss2024smartwatch,anderson2021dementia} and detect epileptic seizures to prevent the development of severe complications~\cite{AmengualGual2019,Shegog2020}.

Multiple attempts have been made to create predictive algorithms to detect AA in PwD using several physiological parameters and/or environmental data~\cite{Spasojevic2021, Khan2017}. Such studies incorporate wearable sensors to capture patient data and use it in AA prediction using machine learning algorithms. Moreover, AI has also been employed in video-based monitoring systems to monitor and detect AA behavior in PwD~\cite{mishra2023privacy, khan2022unsupervised}. To the best of our knowledge, there is no current video surveillance system operating in a hospital setting to detect AA in PwD in real-time due to privacy concerns. The use of multimodal sensing, including wearable sensors and camera footage, for real-time detection of AA and pre-agitation behavior in PwD has not been extensively explored to date. The combination of multimodal sensing and artificial intelligence holds great promise in effectively detecting and predicting AA in real-time. This approach could lead to the timely implementation of preventive strategies or therapies, which could reduce care costs and decrease the frequency of critical incidents among this demographic~\cite{schnaider2002cost,  Canada2022}.

This study aims to understand the complicated behaviors of PwD and predict AA in PwD. We carried out this study in the Geriatric Dementia Unit (GDU) and the Geriatric Transitional Unit (GTU) at the Ontario Shores Center for Mental Health Sciences~\cite{ontarioshores2022} for 5 years. We integrate a multimodal approach, combining biometric data from the EmbracePlus wristband~\cite{EmbracePlus2022} and video data from CCTV cameras installed in common areas in both units. These biometric signs are analyzed to determine the possible correlation with abnormal behaviors. Data collected by these devices, along with the results of the data analysis, is compared against the nurse notes collected via custom forms to confirm the AA and pre-agitation events. The cameras deployed in the designated places automatically detect AA behavior. The developed AI model detects abnormal behavior from body activity recognition in real-time using deep learning techniques. The cameras allow us to document the exact time of the incident for further analysis. The data and analysis then determine personalized pre-agitation conditions using our proposed classification system.

To assess our system, we conducted a pilot study focusing on patient acceptance of wristbands, complemented by video camera validation and multimodal sensor data for predicting AA in PwD. The EmbracePlus wristband was crucial for collecting physiological signals like Electrodermal Activity (EDA), heart rate, skin temperature, and movement data. The camera system was also a key component in detecting body movements and recording agitation events. Both systems are tested on three participants who were successfully recruited at the Ontario Shores Centre for Mental Health Sciences.  We achieved high accuracy in detecting AA through comprehensive data preprocessing, feature extraction, and the ExtraTrees classification algorithm. Additionally, AA detection was enhanced by analyzing real-time video feeds with OpenPose-generated skeletal keypoints and employing RNN-based neural networks, particularly LSTM and GRU~\cite{Carrarini2021AgitationDementia}. These networks, optimized for real-time processing, facilitate timely interventions. The pilot study demonstrated the system's effectiveness through both the wristband and video detection. 

\section{Related Work}
The growing number of PwD causes significant challenges for healthcare systems and caregivers. One of these challenges is to deal with symptoms of AA that increase with the severity of dementia. These symptoms can cause distress, decreased quality of life, and increased healthcare costs. In recent years, there has been a growing interest in developing monitoring systems, that leverage multimodal sensing technologies such as cameras and wearable sensors, to understand and manage AA in PwD~\cite{husebo2020sensing,yang2021multimodal,khan2018detecting}. Wearable wristbands have earned attention as a promising option for monitoring AA in PwD, as some research found a relationship between vital signals, dementia, and aggressive behavior ~\cite{valembois2015wrist,etcher2012non}. These wristbands can capture physiological data, such as acceleration, heart rate, skin conductance, and body temperature, to provide real-time information on PwD AA levels~\cite{badawi2023investigating,badawi2023artificial,Khan2017}. Recent advancements in wearable sensor technology and artificial intelligence have shown promise in addressing the early detection of AA behavior when focusing on signal processing and machine learning to extract features and classify AA events~\cite{cheung2022wrist}.

On the other hand, real-time video-based monitoring systems to monitor AA behavior in dementia patients are an area of interest for researchers today~\cite{khan2018detecting,husebo2020sensing,chikhaoui2016ensemble}. 
These systems use cameras positioned in patients' rooms or common areas to consistently record and monitor patients' behavioral data. Some research ~\cite{mishra2023privacy,khan2022unsupervised} used video cameras to detect AA from previously recorded videos at The Specialized Dementia Unit, Toronto Rehabilitation Institute, Canada. Their system focused on offline AA detection. Another work collected the training dataset from healthy people who imitated agitated hand movements~\cite{marshall2022video}. Researchers have highlighted the importance of real-time feedback, which allows timely interventions and reduces the severity and duration of AA events. This can help improve the quality of life and reduce the stress caused by agitated behaviors in both patients and caregivers. Researchers have also considered privacy factors and concerns while ensuring the accuracy of these systems. The work done in~\cite{mishra2023privacy,marshall2022video} has proposed different masking methods for the people in the video frames that allow for AA detection while keeping the patients' identities and features hidden. 

Multimodal sensing, which combines data from multiple sensors, types, and sources, has emerged as a promising approach to understanding AA in PwD~\cite{rose2015correlates,gong2015home}. It enhances the detection of early signs of AA and the identification of relevant triggers~\cite{chikhaoui2016ensemble,Khan2017}. Clinical trials evaluating the use of multimodal sensing in the context of dementia and AA behavior are crucial in this type of research. These trials assess the feasibility and efficacy of monitoring systems that combine various data sources to inform clinical decision-making. The results of these trials will provide valuable insights into the practical implications of multimodal sensing in real-world healthcare settings. Most existing studies on video systems and wearable sensors for AA detection in real-time have been conducted in controlled laboratory settings or residential care facilities with limited datasets~\cite{yang2021multimodal,ye2019challenges}. Research is needed to utilize more diverse and general datasets gathered and applied in real hospital settings. This is essential since the data from hospitals would be more representative of people with severe dementia who are more prone to AA. The research in such a real-world setting with data gathered from real patients is vital for developing realistic and applicable solutions, ensuring effective treatment and care for PwD~\cite{yang2021multimodal,ye2019challenges}.

Moreover, there is a challenge to identify the actual start and end times of AA episodes. This is due to the fact that the AA events are extracted from nurse notes, which are not accurate and prone to human or individual error. Many AA episodes may also be overlooked and mislabeled as non-agitation. In addition to the datasets, there is a high demand for accurate and reliable end-to-end real-time monitoring solutions to actively predict and respond to AA events. There is also still a need for an in-depth investigation of AI and its features in addition to an in-depth investigation of pre-agitation patterns and signs that could trigger AA in PwD. These investigations are necessary to indicate the usefulness of pre-agitation signs and patterns in early prediction. Digital biomarkers can help detect and predict AA early in real-time~\cite{zhang2022evaluation}. The investigation in~\cite{alam2017motion} shows the correlation between motion biomarkers collected from the accelerometer and the early AA signs, which is particularly useful for personalized AA detection models. 

This paper presents a unique system that combines video analysis and wearable sensor data to predict AA in PwD. The video feeds are crucial for identifying the precise start and end of each AA episode. This precise timing enables us to incorporate data analysis from both the wristband and video footage into our research. We focus on AI and advanced feature engineering to improve the detection accuracy of AA in PwD. We significantly enhance our chances of detecting AA by employing two distinct, yet cooperative models (one based on physiological data and the other on video analysis). This integrated approach also opens the doors to incorporating additional predictive methods, such as audio-based AA detection. The combined use of video-based systems and wearable wristbands offers deeper insights into AA management. While notable advancements have been achieved, further enhancements are needed, especially in real-time accuracy and system applicability. We are confident that our current research explores a vital area and will contribute new knowledge to the field and lay a solid foundation for future advancements.

\section{Methodology}

\subsection{Study Design} 

This study collects participant data using an EmbracePlus wristband~\cite{EmbracePlus2022} and video cameras. The objective of the study is to recruit 20-25 participants in the future from the  Geriatric Dementia Unit (GDU) and the Geriatric Transitional Unit (GTU) at the Ontario Shores Center for Mental Health Sciences. We installed a single AXIS M3077-PLVE Network Camera~\cite{axis_m3077plve} in each unit and one AXIS P3225-VE Mk II~\cite{axis_p3225vemkii_support} in the hallway of GDU. The Research Ethics Board (REB) committee approved this study, ensuring the protection of the participants' privacy and the adherence to the hospital guidelines. We recruit patients based on inclusion criteria such as age, 60 years or older, having a moderately and severe major neurocognitive disorder as defined by a Mini‐Mental State Examination (MMSE) score~\cite{american2013diagnostic}, and being able to ambulate independently (i.e., without the assistance of another person) with or without a walking aid (e.g., walker, cane). We also ensured that the participant has moderately severe major neurocognitive disorder, as defined by a Functional Assessment Staging Tool (FAST) scale score between 6a-6e~\cite{Sclan1992FASTAlzheimers}, has significant agitation or aggression, and meets criteria for agitation in dementia as defined using the 2014 consensus criteria produced by the Agitation Definition Working Group from the International Psychogeriatric Association~\cite{Cummings2015AgitationCognitiveDisorders}.

\begin{figure}[!h]
 \centering
  \subfloat{\includegraphics[width=15cm, height=8cm]{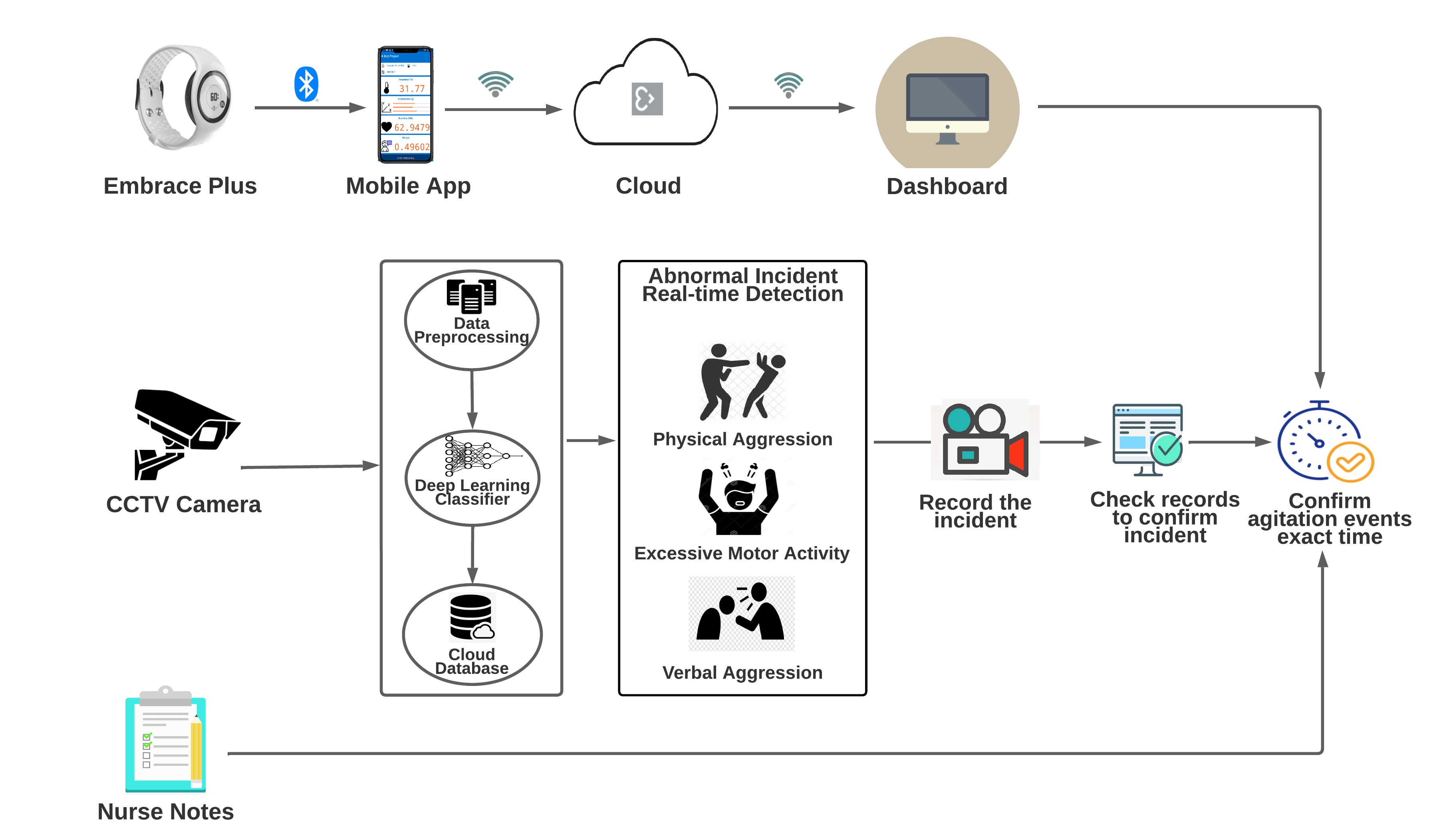}}
 \caption{The workflow of the data collection process.}
 \label{fig:system}
\end{figure}

We follow specific procedures with all patients recruited in this study. The patients enrolled in the study need to wear a wristband during data collection. We closely monitored for signs of discomfort when wearing the wristband. In case the patient refuses to wear the wristband, a nurse re-approaches and attempts to apply the wristband later. The participants wear the device for 24 to 72 hours on three separate occasions within the six-week study period. The EmbracePlus device collects participants' physiological parameters. We record the baseline measurements of vital signs during the initial visit as the referring data points. Clinical staff monitor participants, make notes of any episodes of AA, and include the start and end times of the events. As there can be a lag between the actual episode of AA and the time a nurse observes and records the episode, the CCTV cameras in the units record video footage of the participants. These cameras capture real-time footage of the participants’ episodes of AA. We mask data collected by these cameras and any faces appearing in the video frames are blurred. The video detection system processes the data, classifies AA, and records the incident to be checked by the healthcare provider. The exact time of the AA incident and comprehensive incident details are documented in the camera footage. The physiological parameters from the three sources provided are analyzed to identify changes during and before episodes of AA in PwD. This study is currently ongoing at Ontario Shores Mental Health Institute~\cite{Choudhury_Badawi}. We have successfully recruited three participants using the established procedures, and we present the preliminary results of this work. The system architecture of data collection is shown in Figure \ref{fig:system}.

\subsection{Addressing Issues Arising from the REB Approval}
The research commenced in 2019, with an REB approval process that lasted about six months. A significant challenge was presented by the use of video cameras to document the behaviors and activities of patients, staff and visitors in the public spaces of hospital inpatient units. 
Consent was secured for these patients through substitute decision makers (SDM), in light of their advanced dementia condition. At admission, capacity assessments were performed and SDMs were approached for permission to involve patients in potential research. 
Following the receipt of ethics approval, the research team organized informational sessions within the inpatient unit and restricted video recording to only the common areas where patients usually gather during the day. Recording devices were intentionally not placed in private areas such as patient rooms, restrooms, staff meeting rooms, and nursing stations, and audio capture was turned off throughout the data collection period. To alleviate the concerns about the REB's privacy and confidentiality, stringent data protection measures were instituted, including encrypting and securing all data on hospital servers with password protection. Clinical staff were tasked with observing participants for any agitated actions (AA) as part of their regular duties. 

\subsection{Event Classification}
The proposed system collects the biomarkers using the EmbracePlus wristband, which is considered the state-of-the-art wearable device for continuous health monitoring in the market today~\cite{EmbracePlus2022}. The device combines digital biomarkers, robust sensors, and a user-friendly design to continuously monitor participants with various health conditions. It collects Electrodermal (EDA) that detects slight changes in skin conductance from the skin surface, Photoplethysmogram (PPG) that calculates the pulse rate and pulse rate variability measurements, skin temperature, and raw accelerometry data for motion detection. The collected signals are sent to the cloud-based EmbracePlus Care platform~\cite{EmbracePlus2022} through a Bluetooth-connected gateway (e.g., a smartphone).

\begin{figure*}[!h]
 \centering
  \subfloat{\includegraphics[width=15cm]{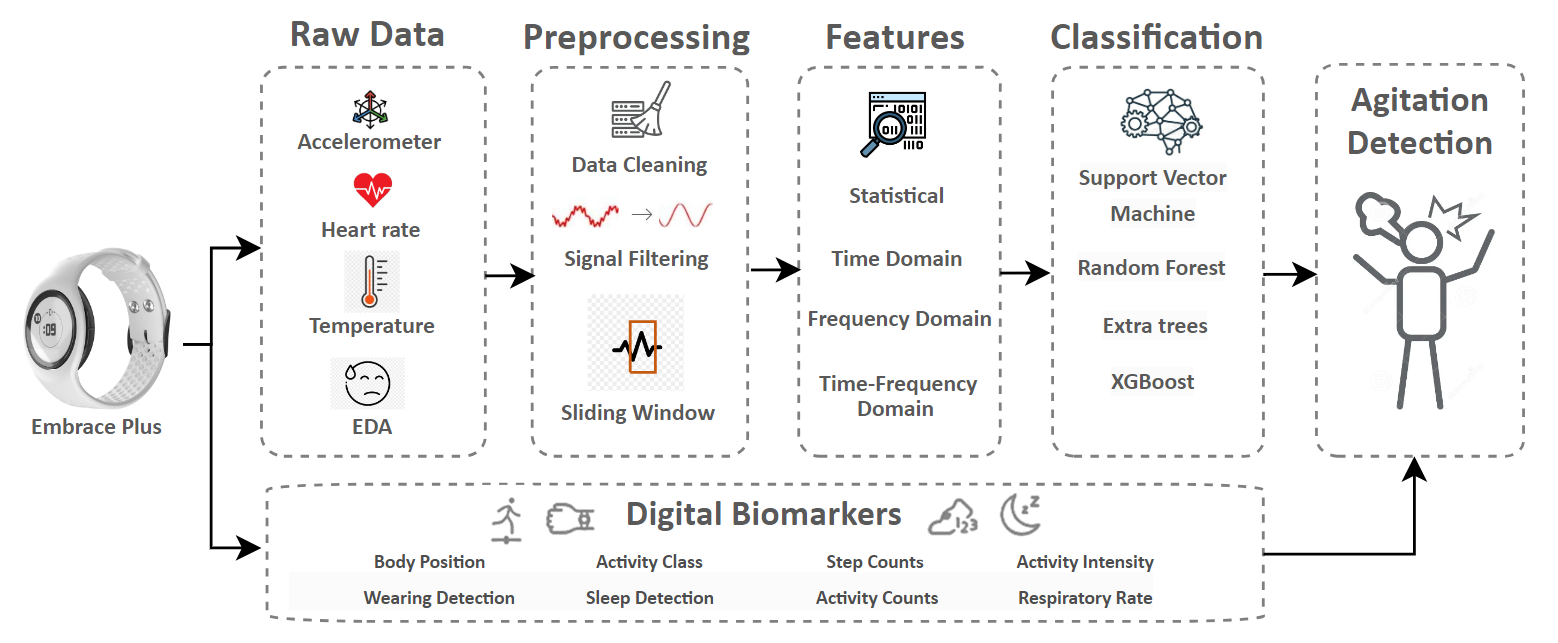}}
 \caption{The Workflow of the Proposed Classification System using the EmbracePlus Wristband.}
 \label{fig:ml}
\end{figure*}

The first type of data we deal with is the raw data from the accelerometer, heart rate, temperature, and EDA signals. We follow several pre-processing steps to clean, filter, and apply one-minute window segmentation for the raw signals~\cite{badawi2023investigating,badawi2023artificial}. We then extract features from the signals as shown in our previous work from the statistical, time domain, frequency domain, and time-frequency domain with around 150 features~\cite{badawi2023investigating,badawi2023artificial}. Lastly, we evaluate multiple classification techniques, namely Support Vector Machine (SVM), Random Forest, Extra Trees, and Gradient Boosting to classify AA events. The performance of each model is evaluated using standard classification metrics such as accuracy, precision, recall, AUC, and F1-score. Furthermore, we collect the digital biomarkers that are pre-processed data derived from Empatica's algorithms and calculated minute-by-minute. The second type is the digital biomarkers, which include Pulse Rate Variability, Respiratory Rate, Movement Intensity, Accelerometer Magnitude Standard Deviation, Steps, Skin Conductance Level (SCL), Wearing Detection, Temperature, and Sleep Detection as shown in Table \ref{tab:my-table}. Digital biomarkers have the capability to effectively and accurately oversee human health from a distance, consistently, and without causing disruption. This applies across a spectrum of health conditions ~\cite{EmbracePlus2022}.  Figure \ref{fig:ml} shows the classification workflow from the EmbracePlus wristband using raw data and digital biomarkers.

The proposed work focuses on investigating the data from PwD using machine learning, two of which were thoroughly investigated in our previous work~\cite{badawi2023investigating,badawi2023artificial}. The results concluded the most important features for this problem after performing feature engineering and proved that personalized models on individual patients outperform generic models. In this work, we report the results of the personalized model on three different participants from the Ontario Shores Mental Health Institute. 
We test our system in real-time once we determine the optimal classification system to predict AA events. In the real-time (online) detection phase, real-time raw data is transmitted from the wristband. Following this, features are extracted from each 1-minute window, and these specific features are fed into the customized model to classify whether the data is considered normal or indicative of AA. The outcomes are subsequently transmitted to the backend system, and the healthcare provider is notified if the patient is agitated. 

\begin{table}[]
\caption{The Digital Biomarkers Data Description from the EmbracePlus Wristband.}
\label{tab:my-table}
\centering
\resizebox{\textwidth}{!}{%
\begin{tabular}{|c|c|}
\hline
Digital Biomarkers           & Definition                                                                                                        \\ \hline
Pulse Rate (PR)              & Algorithm utilize PPG and accelerometer data for PR monitoring with estimates on 10-second windows.            \\ 
Pulse Rate Variability (PRV) &
  Algorithm analyzes PPG for intermittent PRV, using accelerometer signals with non-overlapping windows. \\ 
Respiratory Rate (RR)        & Algorithm processes PPG and accelerometer data to calculate RR values expressed in breaths per minute       \\ 
Movement Intensity           & Algorithm calculates activity count, steps, and accelerometer std from the accelerometer sensor. \\
Skin Conductance Level (SCL) & SCL estimation from EmbracePlus EDA signal, output every 1 minute with non-overlapping windows.                             \\ 
Wearing Detection            & Algorithm correlates device status with PPG patterns, indicating wearing time.                                              \\ 
Temperature                  & Algorithm analyzes EmbracePlus data for continuous peripheral temperature estimation.                                       \\  \hline
\end{tabular}%
}
\end{table}

\subsection{Video-based Analysis}
In addition to collecting data on physiological biomarkers, the study incorporates video analysis data for AA prediction. This approach utilizes an extra cooperative model, which improves our overall AA detection system and allows us to get the precise duration, including start and end times, of the collected AA episodes. Moreover, once an AA episode is detected, the cameras record a previously set pre-agitation, making it easier to observe any visual pre-agitation signs. We aim to provide real-time alerts to healthcare providers for timely intervention. The setup includes three CCTV cameras installed and a PC in the attending psychiatric office with access to this footage. Our system operates in two phases: the offline phase for manual labeling and model training, and the real-time stage for running the model. To protect the privacy of the participants and the staff present, we blur all faces and run OpenPose, a computer vision tool that extracts skeletal keypoints, to capture movement data~\cite{8765346,simon2017hand,cao2017realtime,wei2016cpm}. The model is trained on features extracted from keypose points instead of the raw video frames. This approach has been recently used by researchers for AA detection in PwD, and has proven to be as successful in detecting AA while preserving the privacy of the people present~\cite{mishra2023privacy,marshall2022video}. 

In our research, we utilize OpenPose, which is an advanced real-time system for multi-person 2D pose estimation, to anonymize individuals in video frames. OpenPose utilizes Convolutional Neural Networks (CNN) to detect human body parts and map their skeletal structure onto the image or video frame. This allows for a detailed representation of movement data present in the collected frames. After this, we employ a preprocessing phase to enhance the generalizability of the model across various environments and datasets. This phase involves the elimination of extraneous noise that could otherwise impede model performance. The model considers the variations in camera angles and subject positioning within the frame, which can significantly influence the coordinate data. We calculate Euclidean distances and angle measurements between specific skeletal coordinates to determine movements. For example, the measured distance between the torso and feet is useful to identify potential kicking actions, which may indicate AA in certain contexts. Table \ref{tab:table2} summarizes all the 47 features extracted from these distance and angle measurements. Before training, a feature analysis step is introduced to remove highly correlated features and reduce the dimensionality of the model. This process reduced the features to 15 features as shown in Table []. We tested the system on the same three participants whose wearable sensor data was used earlier.

\begin{table}[h]
    \caption{Dataset Description.}    
    \centering
    \label{tab:table2}
    \resizebox{\textwidth}{!}{%
    \begin{tabular}{|c|c|}
            \hline
            Feature & Description \\  
            \hline
            eu\_1-eu\_14 & \makecell{Euclidean distances between different keypoint pairs eu\_1 represents the Euclidean \\distance between keypoint 1 and the previous position of keypoint 1} \\ 
            eu\_1\_3-eu\_1\_14 & Euclidean distances between keypoint 1 and various other keypoints \\
            
            por\_2\_1-por\_14\_1 & Point of reference (POR) values between keypoints 2-14 and keypoint 1  \\
            ang\_1\_2-ang\_1\_14 & Angles between keypoint 1 and keypoints 2-14  \\
            
            \hline
\end{tabular}%
}
\end{table}
            
\begin{figure}[h!]
 \centering
 \subfloat{\includegraphics[width=13cm]{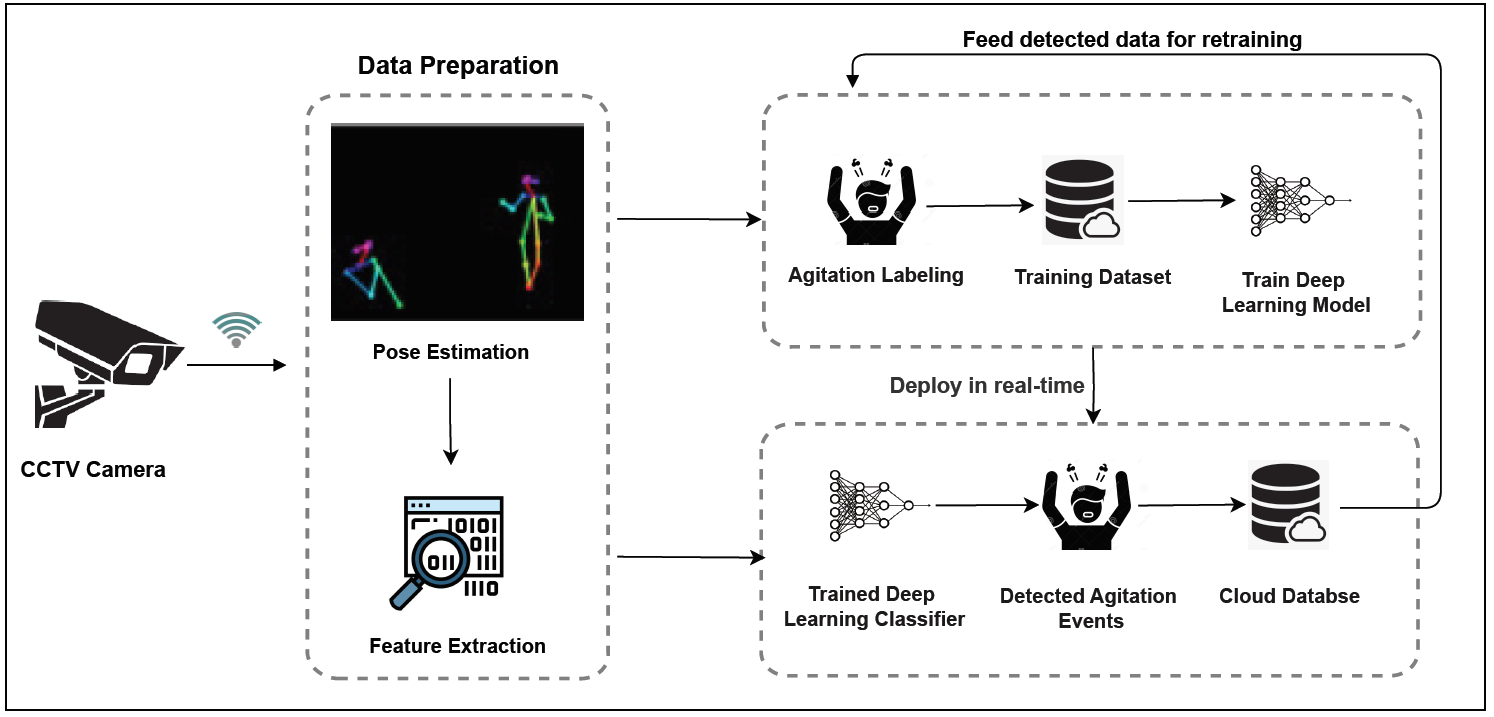}}
 \caption{The System Architecture of the Video-based Detection.}
  \label{fig:video}
\end{figure}

The offline and real-time (online) stages of the system are detailed in Figure \ref{fig:video}. In the offline stage, we preprocess and extract features from skeletal data, which are then labeled using nurse notes from patient medical records. Access to the collected videos is restricted to the computer in the psychiatric office and they are retained only until the AA episodes are accurately labeled with their start and end times. Once the dataset is finalized, all videos are securely discarded. Using this dataset, we train a deep learning model to differentiate between AA and non-agitation events. Our focus is on capturing a range of behaviors, from violent or aggressive actions to repetitive motions like pacing or chair rocking. Hence, we use models that consider sequences of actions, such as RNN-based neural networks, to effectively recognize these sequences of actions.
We specifically utilize and compare the results of the GRU (Gated Recurrent Unit) and LSTM (Long Short-Term Memory) models. Both are adept at analyzing sequences of actions. The LSTM model is designed to capture long-term dependencies within sequences. The sophisticated cell structure of LSTM cells makes it highly effective in maintaining context over long intervals. However, the GRU model employs a simpler architecture that aims to achieve results comparable to LSTM models, but with lower computational costs. Both models utilized in our research are composed of a single LSTM or GRU cell, followed by a fully connected sigmoid layer for the binary classification of AA episodes.

In the real-time stage, we deploy our offline trained model to classify AA in real-time. The model first processes real-time video data from hospital cameras. Video frames are analyzed using OpenPose, extracting similar features as in the offline stage. We use a fixed-size window to input frame sequences into the classifier. Upon detecting AA, the system records the event, including a five-minute buffer before and after the incident to capture the entire context. This approach helps identify potential triggers and patterns that lead to AA. The psychiatrist reviews these detected events for accuracy and confirms AA events are added to our training dataset with appropriate labels. The model is then retrained in the offline stage with the added data. The primary goal of retraining the model is to continuously adapt and improve the model with newly detected data.

\section{Preliminary Data Analysis and Evaluation}
A pilot study was conducted to validate the effectiveness and feasibility of the proposed system at Ontario Shores Mental Health Hospital. This initial investigation aimed to provide valuable insights into the system's functionality, usability, and overall potential before the implementation on a larger population. Details of the enrolled participants and the data collected can be found in Table \ref{tab:tablex}. Upon enrolment, the participants wore the EmbracePlus wristband on three different days. We turned on the cameras installed in the unit during the data collection days to record the participants' activities. Lastly, we assigned a nurse to observe the participant and provide a detailed report of behavior, AA events, and any abnormal behavior. During the three days, we collected six AA events ranging from two to twenty-three minutes per AA event with a total of 20-32 minutes of AA labels and 560-581 minutes of normal labels for each participants. The following sections will present the results in detail from the EmbracePlus wristband and video cameras.

\begin{table}[h!]
    \caption{Overview of collected Data.}    
    \centering
    \label{tab:tablex}
    
    \begin{tabular}{|c|c|c|c|}
            \hline
            Participant  & Gender & Age & Total Collected Data (hours)\\
\hline
1 & F & 83  & 54.21 \\
2 & F & 63 & 48.13 \\
3 & M & 67 & 36 \\

            \hline
\end{tabular}%

\end{table}

\subsection{Performance Evaluation}
\subsubsection{The EmbracePlus Wristband Raw Data}

This study utilize raw data obtained from the wristband four signals and used personalized models, which achieved superior accuracy in AA detection from PwD in previous research~\cite{badawi2023investigating,badawi2023artificial}.
Subsequently, we conducted a comparative analysis of multiple machine learning algorithms for AA detection, including Support Vector Machines (SVM), Random Forest, Extra Trees, and Gradient Boosting. We trained and tested a personalized model for every participant and reported the evaluation results in Table \ref{tab:wearable_comp}. The dataset for each participant was randomly split into 70\% training and 30\% for testing.

The Extra Trees model emerged as the top-performing algorithm for all three participants. For participant \#1, the Extra Trees model achieved an accuracy of 98.67\%, an AUC of 99.1\%, a recall of 99.76\%, and an F1-score of 98.70\%. It achieved the highest accuracy and AUC for participant \#2 of 90\% and 98\%, respectively. Similarly, for participant \#3, it achieved the highest accuracy of 99\%. These results underscore the efficacy of the chosen features, preprocessing methodologies, and up-sampling techniques. While acknowledging the potential concern of overfitting, collecting additional patient data over time is anticipated to solve this issue. All models were also tested on the three participants together. The Extra Trees model achieved a much lower accuracy of 69\%. The XGBoost and Random forest achieved higher accuracies in comparison with 98\% and 93\%, respectively. 

\begin{figure}[h!]
\begin{subfigure}{\textwidth}
    \centering
    \includegraphics[scale=0.6]{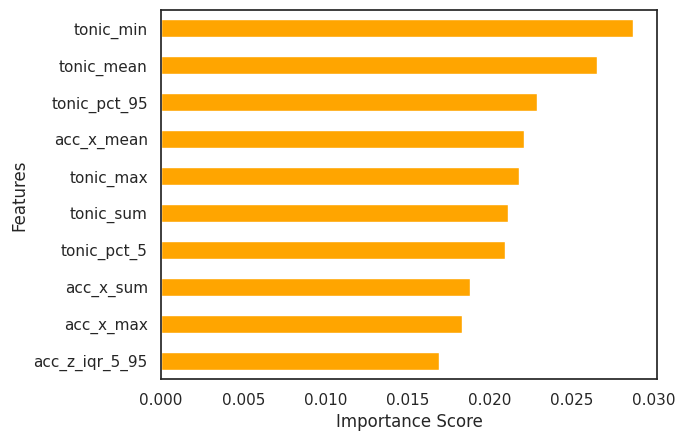}
    \caption{Feature Importance for Participant \#1.}
    \label{fig:p1f}
\end{subfigure}
\begin{subfigure}{\textwidth}
    \centering
    \includegraphics[scale=0.6]{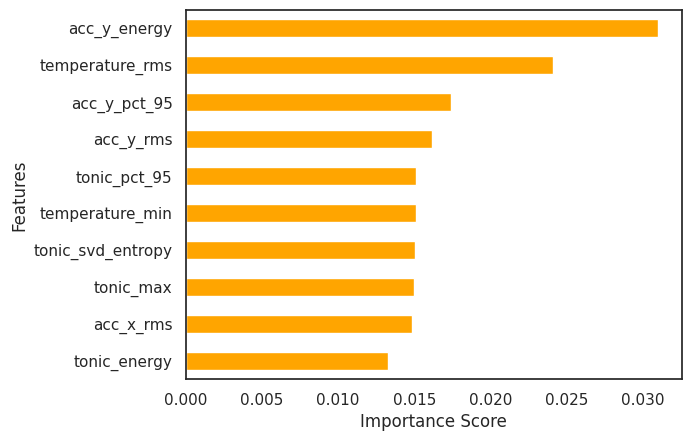}
    \caption{Feature Importance for Participant \#2.}
    \label{fig:p2f}
\end{subfigure}

\begin{subfigure}{\textwidth}
    \centering
    \includegraphics[scale=0.6]{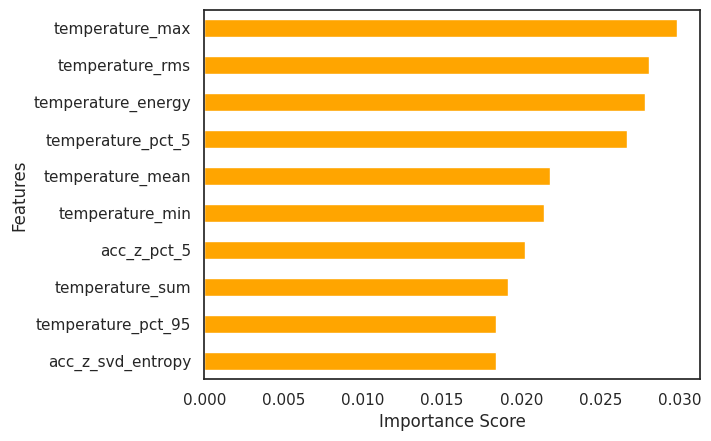}
    \caption{Feature Importance for Participant \#3.}
    \label{fig:p3f}
\end{subfigure}
        
\caption{The Feature Importance Plots for the Participants.}
\label{fig:figures}
\end{figure}

\begin{figure}[h!]

\begin{subfigure}{\textwidth}
    \centering
    \includegraphics[width=0.9\textwidth]{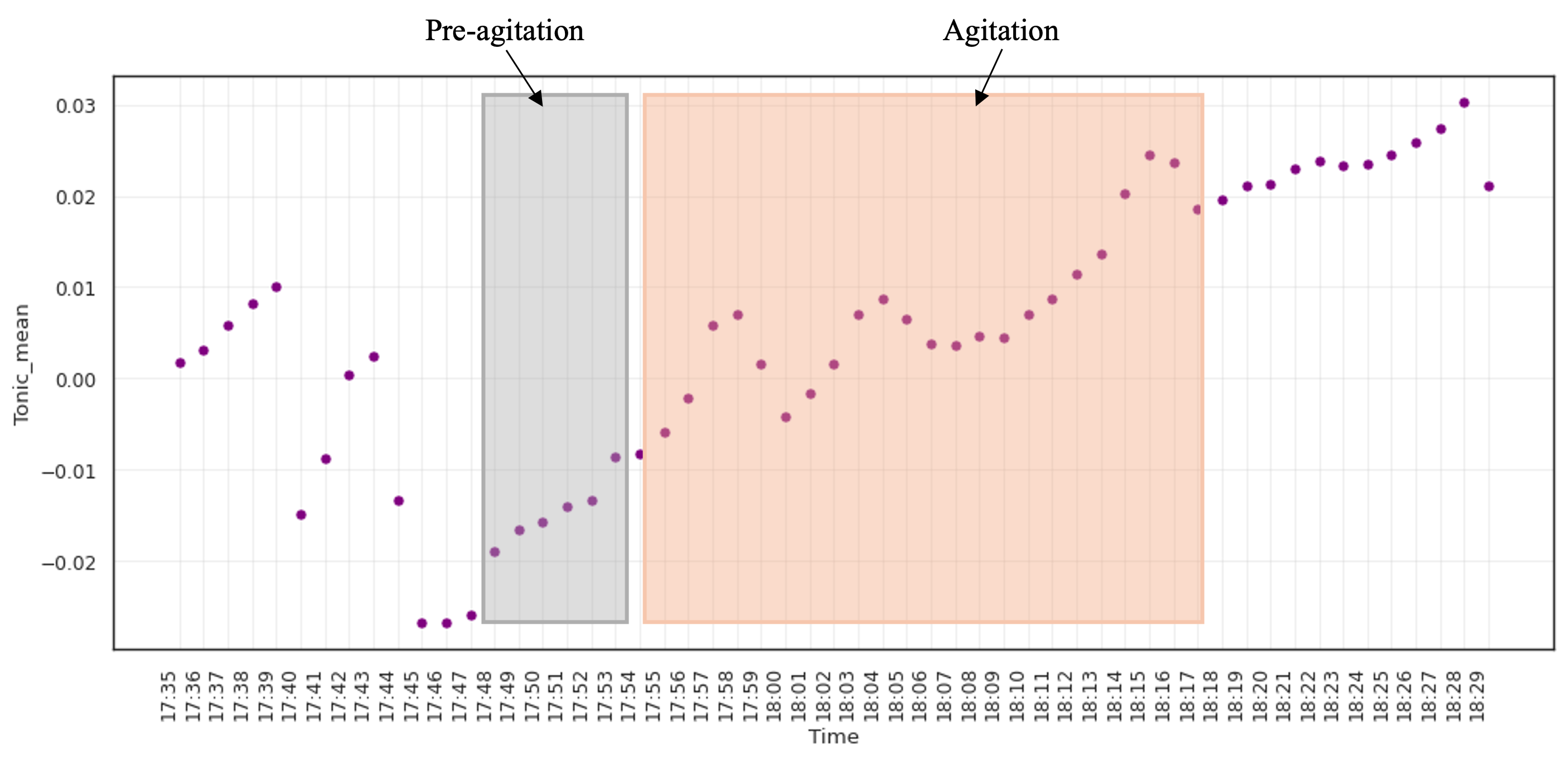}
    \caption{The Tonic Mean Plot for Participant \#1. }
    \label{fig:p1}
\end{subfigure}

\begin{subfigure}{\textwidth}
    \centering
    \includegraphics[width=0.9\textwidth]{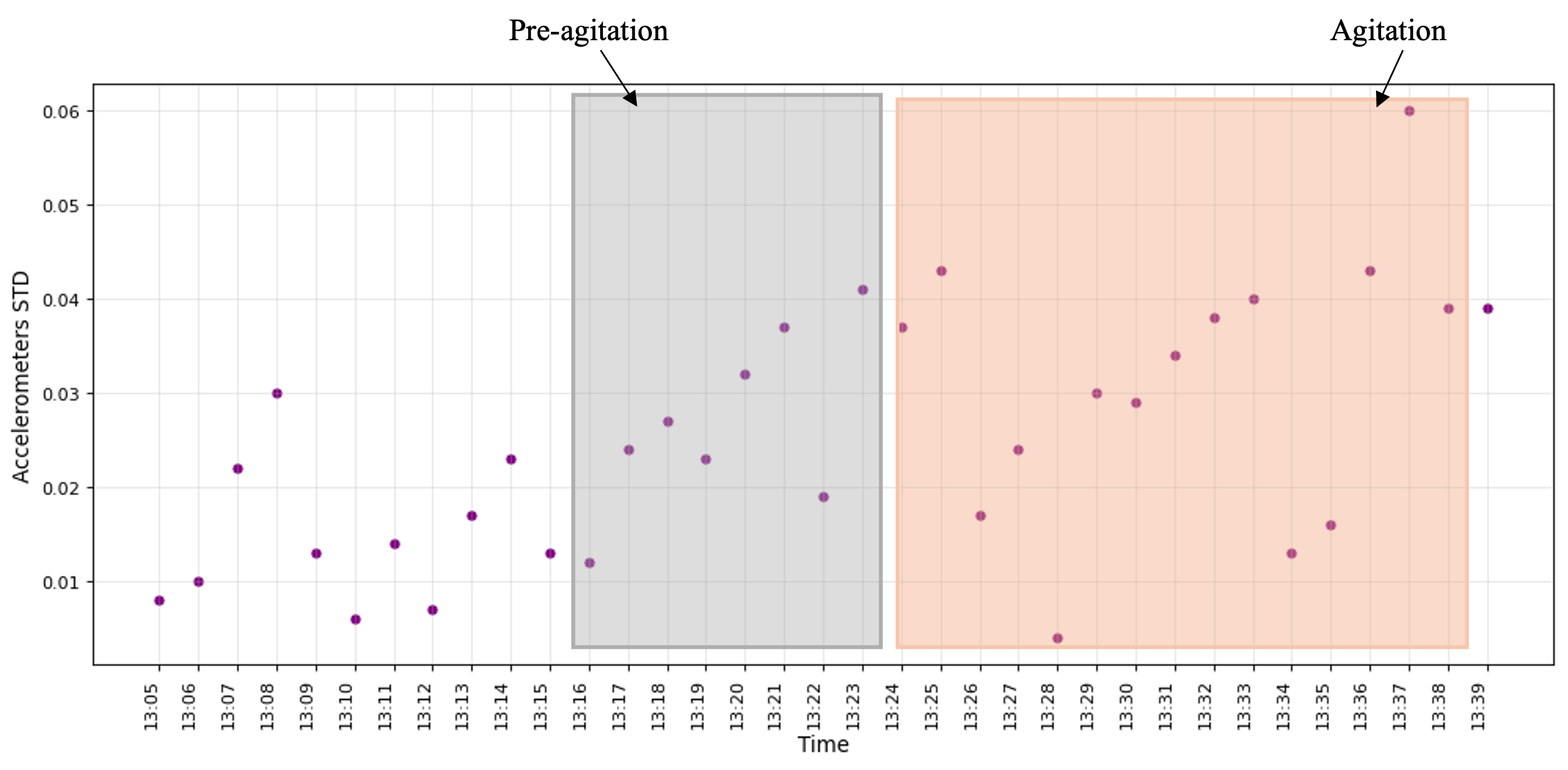}
    \caption{The Accelerometer Plot for Participant \#2.}
    \label{fig:p2}
    \begin{subfigure}{\textwidth}
    \centering
    \includegraphics[width=0.9\textwidth]{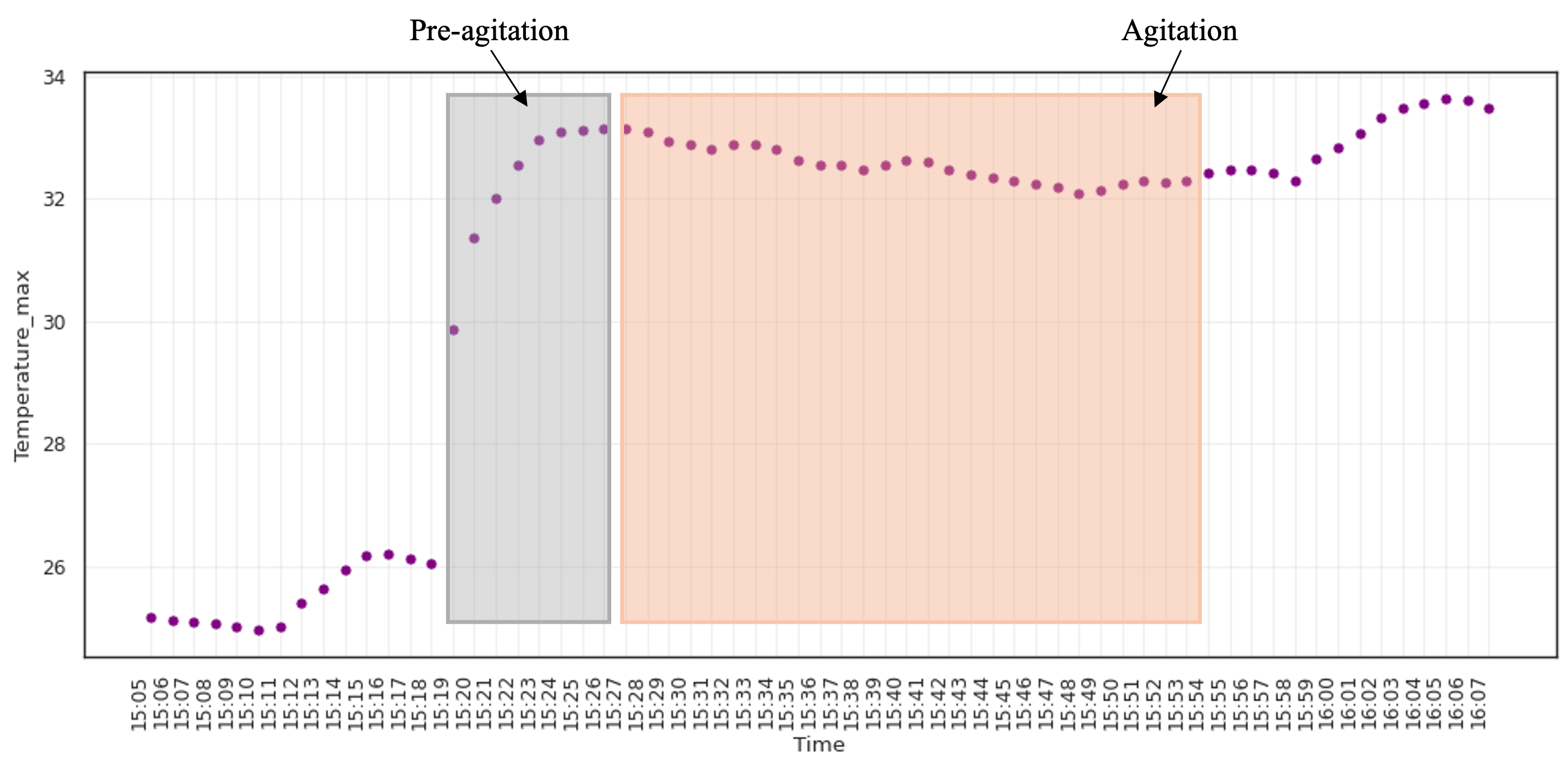}
    \caption{The Temperature Plot for Participant \#3.}
    \label{fig:p3}
\end{subfigure}
\end{subfigure}
        
\caption{The EmbracePlus Wristband raw data with agitation and pre-agitation annotations.}
\label{fig:raw_figures}
\end{figure}

The Random Forest model, on the other hand, performed very poorly in other evaluation matrices with an F1-score of 51\% and a recall of 37\%. The XGBoost performed best on all the participants, which highlights the potential for a general model when enough data is collected. Furthermore, for participant \#1, the top 10 features contributing to accurate AA classification using the Extra trees model revealed that five were from EDA, three from the accelerometer, one from heart rate, and one from temperature. Figure \ref{fig:p1f} shows a summary of the feature importance plot for this participant. Since the EDA tonic mean was the top feature to classify AA, we investigated in-depth the AA labels. Figure \ref{fig:p1} shows the Tonic mean values of the first participant from the EDA signal during labeled AA events from the camera and nurse notes (highlighted in red). This event occurred during the second day and lasted for 23 minutes from 17:55 pm to 18:17 pm. This observation suggests that the patient's AA was related to the EDA signal connected to the emotions. We also observed an apparent change to the data before the actual AA occurred, which we manually marked as pre-agitation labels (highlighted in blue).

For participants \#2 and \#3, the feature importance plot changes as shown in figures \ref{fig:p2f} and \ref{fig:p3f}. These plots show that, for these two participants, the features related to acceleration and temperature were the most important in AA detection, respectively. For participant \#2, the acceleration features were the top features in identifying the AA event. Figure \ref{fig:p2} shows one of the AA events for this participant occurring between 13:24 and 13:38 using the accelerometer data. Moreover, a change in the pattern of the signal eight minutes before the event is manually labeled as pre-agitation. For participant \#3, the most dominant features were the temperature features, so an example of the temperature readings for an AA event is shown in Figure \ref{fig:p3}. Just as before, a change of the pattern before the observed AA event is manually labeled as pre-agitation. These observations suggest the potential for detecting pre-agitation patterns from raw data, enabling the prediction of AA before it occurs.
\begin{table}[h]
    \caption{Comparative performance metrics using Raw Data.} 
    \centering
    
   \begin{tabular}{|c|c|c|c|c|c|}
            \hline
            Participant & Model & Accuracy & AUC & F1 score & Recall \\  
            \hline
            \multirow{1}{*}{1} & Extra Trees & \textbf{0.98} & \textbf{0.99} & \textbf{0.98} & \textbf{0.99} \\
                               & XGBoost & 0.97 & 0.98 & 0.97 & 0.98 \\ 
                               & Random Forest & \textbf{0.98} & \textbf{0.99} & \textbf{0.98} & \textbf{0.99} \\ 
                               
            \hline
            \multirow{3}{*}{2} & Extra Trees & \textbf{0.90} & \textbf{0.98} & 0.90 & \textbf{0.96} \\ 
                               & XGBoost & 0.88 & 0.96 & 0.86 & 0.93 \\ 
                               & Random Forest & 0.88 & 0.96 & 0.89 & 0.94 \\ 
            \hline
            \multirow{3}{*}{3} & Extra Trees & \textbf{0.99} & \textbf{0.99} & \textbf{0.99} & \textbf{0.99} \\ 
                               & XGBoost & 0.98 & 0.99 & 0.98 & 0.99 \\ 
                               & Random Forest & 0.98 & 0.99 & 0.98 & 0.98 \\

            \hline
            \multirow{3}{*}{All} & Extra Trees & 0.69 & 0.94 & 0.91 & \textbf{0.93} \\ 
                               & XGBoost & \textbf{0.98} & \textbf{0.96} & \textbf{0.95} & 0.88 \\ 
                               & Random Forest & 0.93 & 0.95 & 0.51 & 0.37 \\

            \hline
    \end{tabular}%
 \label{tab:wearable_comp}
\end{table}

\subsubsection{The Wristband EmbracePlus Digital Biomarkers}
We explored all the digital biomarkers offered by EmbracePlus and observed that pulse rate, activity counts, and activity class were the leading indicators for AA detection for all three participants. In Figure \ref{fig:figures}, the same AA events discussed in the previous subsection for the three participants are illustrated and the values during labeled AA events from the camera and nurse notes are highlighted in red. Additionally, we observed a noticeable change in the data before the onset of AA, manually designated as pre-agitation labels and highlighted in grey. The manual label of the pre-agitation was done after reviewing all the signals for the participants and noting the same change across multiple patterns.

\begin{figure}[h!]

\begin{subfigure}{\textwidth}
    \centering
    \includegraphics[width=0.9\textwidth]{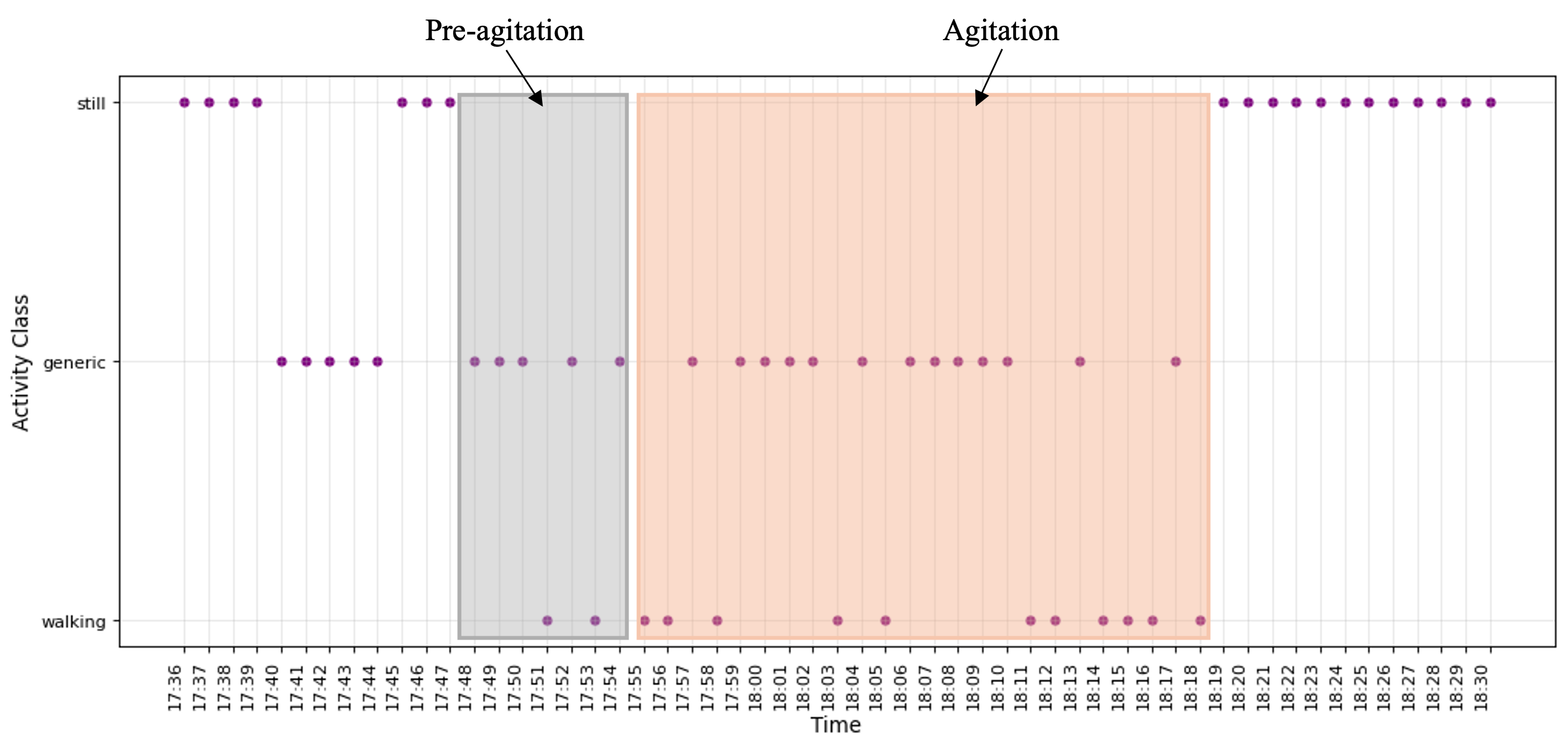}
    \caption{The activity class from the digital biomarkers data for participant \#1.}
    \label{fig:first}
\end{subfigure}

\begin{subfigure}{\textwidth}
    \centering
    \includegraphics[width=0.9\textwidth]{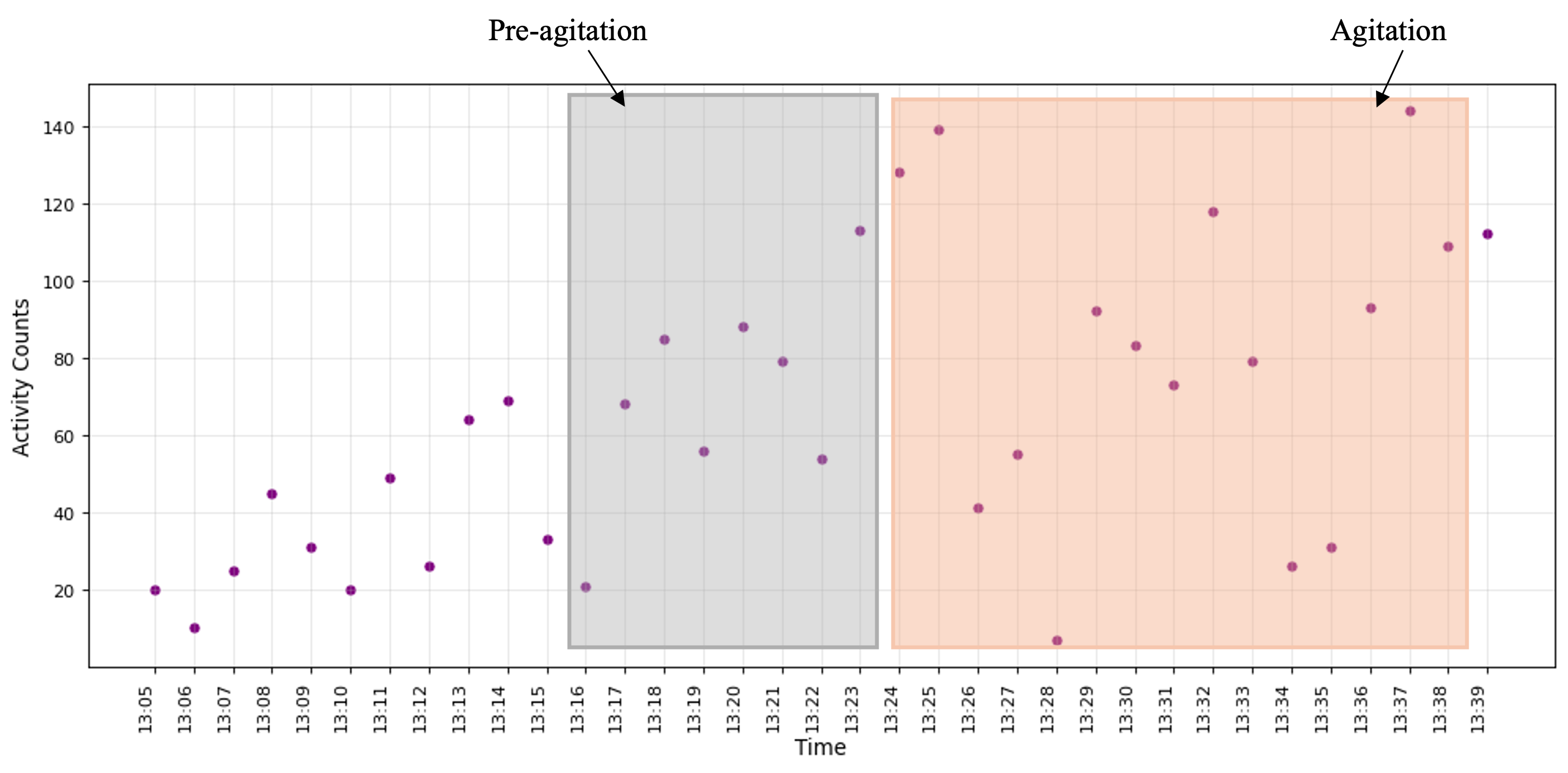}
    \caption{The activity counts from the digital biomarkers data for participant \#2.}
    \label{fig:second}
    \begin{subfigure}{\textwidth}
    \centering
    \includegraphics[width=0.9\textwidth]{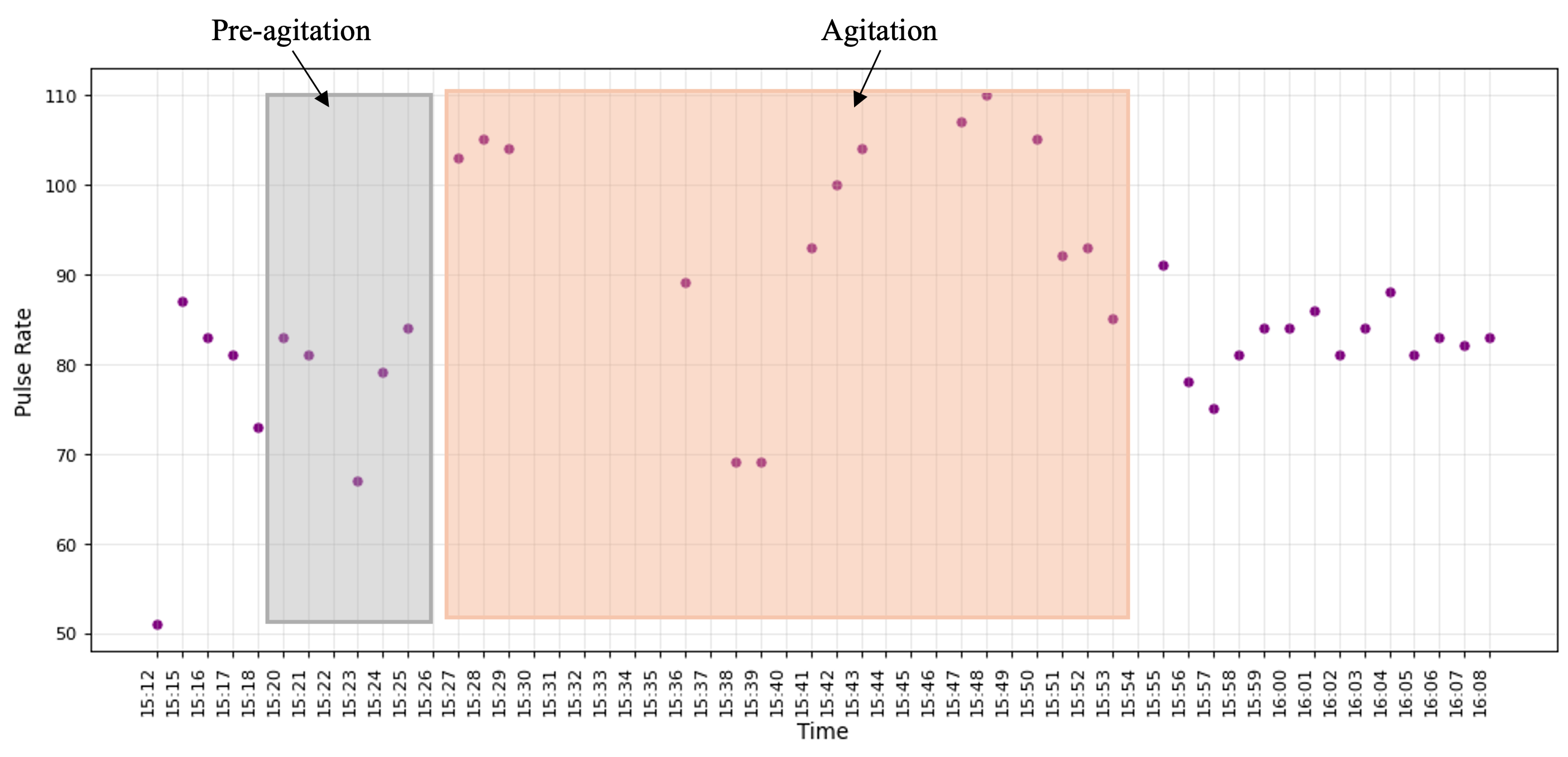}
    \caption{The pulse rate from the digital biomarkers data for participant \#3.}
    \label{fig:third}
\end{subfigure}
\end{subfigure}
        
\caption{The digital biomarkers with agitation and pre-agitation annotations.}
\label{fig:figures}
\end{figure}

Figure \ref{fig:first} illustrates the activity class, for the participant \#1, extracted from the accelerometer signal, revealing that the participant was in motion rather than stationary during AA and pre-agitation episodes, indicating body movement during these events. Figure \ref{fig:second} displays the total activity counts for participant \#2 from the accelerometer signal. While the normal activity count for the participant ranged between 0-100 during AA and pre-agitation events, it surged to 50-140, signifying heightened activity levels during AA. Finally, Figure \ref{fig:third} presents the pulse rate derived from the heart rate signal for participant \#3. Although the participant's average pulse rate ranged from 55-80 bpm, it increased to 90-110 bpm during AA and pre-agitation events. Across the raw and digital biomarkers data, we observed that the pre-agitation occurred from 15:20 to 15:27 pm. This indicates that signs of AA behavior occurred approximately seven minutes before the actual event, suggesting the potential to predict and prevent AA events.

\begin{figure*}[!h]
 \centering
  \subfloat{\includegraphics[scale=0.45]{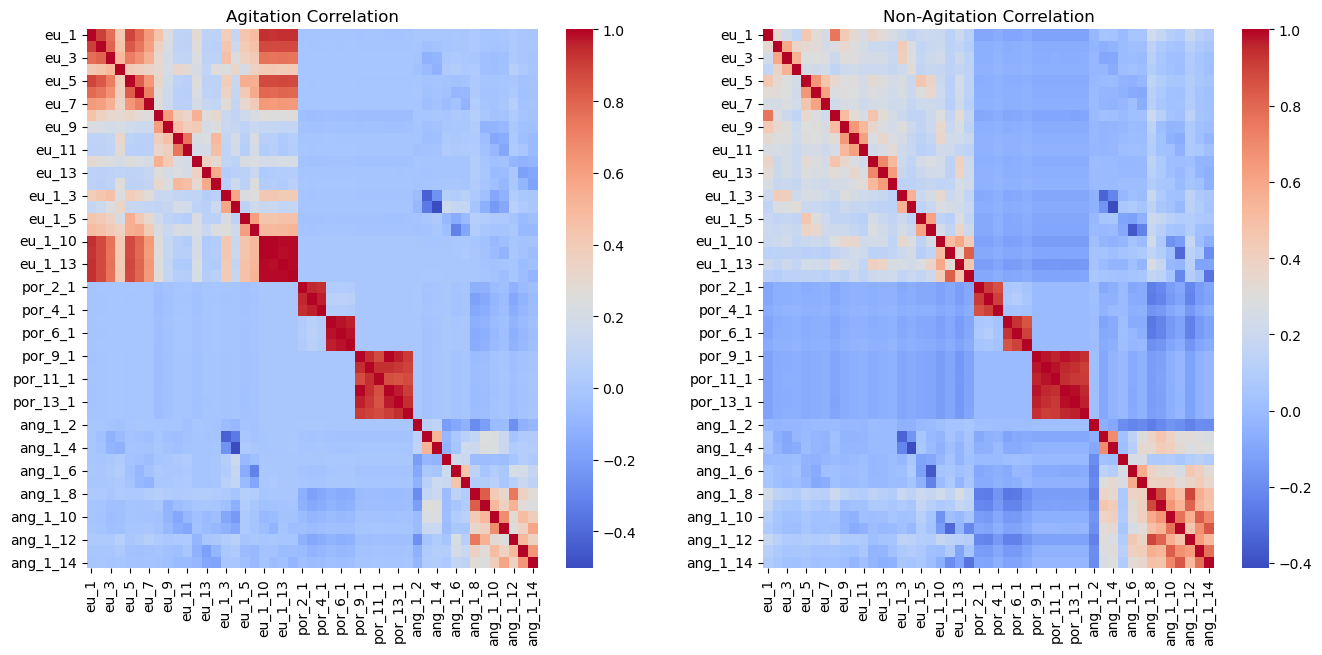}}
 \caption{The correlation of the video features for both AA and non-agitation events.}
 \label{fig:corr}
\end{figure*}

\subsection{Performance of Video-based Detection}
We preprocessed our videos using OpenPose and performed feature extraction as described in the methodology section. Since AA behaviors are repetitive in nature, we selected recurrent neural network (RNN) models to capture the sequential patterns of these events. Features were extracted from 30-second windows. The window moves one second at a time to capture different variations of AA behaviors from the skeletal points. This resulted in 182,994 AA sequences and 184,786 non-agitation sequences. For the classification task, we tested three different network structures that are Long Short-Term Memory (LSTM) model and a Gated Recurrent Unit (GRU). 

In the feature analysis step, we focused on reducing the model's dimensionality without compromising its performance. Initially, 47 features were extracted based on skeletal movements, including Euclidean distances and angles between key body parts. Figure \ref{fig:corr} shows all the 47 features and their correlation. It is evident from the figures that more features are highly correlated in agitation events than in non-agitation events. This had a strong effect in the feature reduction since only the highly correlated features in both datasets were removed. We invistaged the correlation of the features deeper and tested the model on fewer features based on a correlation threshold. Setting the correlation threshold to 0.8 reduced the number of features to 39 features. The number of features based on the correlation threshold did not change even when the threshold was set to as low as 0.5 due to the huge difference in correlation between both types of events. Both datasets are randomly split into 70\% for training and 30\% for testing. The training was conducted on a lambda server equipped with an RTX A6000 GPU, and each model was trained for 100 epochs with a batch size of 256. We employed the Adam optimizer to efficiently handle sparse gradients and used a sigmoid activation function for binary classification (AA vs. non-agitation). We report the results of all the tests on the testing set in Table \ref{tab:comp}.

\begin{figure*}[!h]
 \centering
  \subfloat{\includegraphics[scale=0.8]{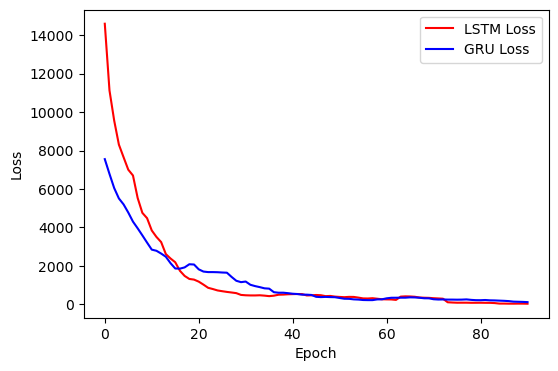}}
 \caption{Loss Comparison between LSTM and GRU Models.}
 \label{fig:loss}
\end{figure*}

We compared metrics such as accuracy, AUC, and F1-score. We also compared the response time, an essential factor in real-time applications, of all models. As observed in the table, the reduction in the number of features did not affect the performance of the models. The LSTM model, in both cases, achieved an accuracy of 98\% and an AUC of 99\%. The GRU model reached 99\% accuracy and 99\% AUC. Although the performance of both models is similar, the response time of the GRU model is double that of the LSTM. The response times for the LSTM was 15.9 seconds when all the features were used in training and was almost one second faster with less features. The GRU models, on the other hand, had a response time of 30.1 seconds for 47 features and 29.7 seconds for 39 features. The results show that the GRU model is superior across all evaluation metrics, albeit for the response time, where it lags behind the LSTM model by a huge margin. Since we aim to detect AA as early as possible, the swifter response time can allow for timely interventions by healthcare providers in case of any AA event. The high AUC values of both models signify a strong ability to minimize the rate of false positives. This is crucial to ensure the reliability of the model in detecting real AA with lower false alarms, causing less overhead for healthcare providers. Moreover, we report the loss curves of both models in Figure \ref{fig:loss}. As shown in the figure, both models demonstrated a smooth stable descent; however, the LSTM model converges faster than the GRU model. This, along with the faster response time, posits the LSTM model as possibly the more advantageous model for real-time deployment.

\begin{table}[h]
    \caption{Video Results Comparison.} 
    \centering
    
    \begin{tabular}{|c|c|c|c|c|c|c|}
            \hline
            Model& Number of features & Accuracy & AUC & F1 score & Recall &Time(s)\\  
            \hline
            LSTM& 47& 0.98& 0.99 &0.98 & 0.96 &15.9\\ 
            LSTM& 39& 0.98& 0.99 &0.98 & 0.97 & \textbf{14.5}\\

            GRU & 47&\textbf{0.99}&\textbf{0.99}& \textbf{0.98}&0.97 & 30.1 \\
            GRU & 39&\textbf{0.99}&\textbf{0.99}& \textbf{0.98}&0.97 & 29.7 \\

            
            \hline
\end{tabular}%
 \label{tab:comp}
\end{table}

\section{Discussion and Future Directions}
The successful implementation of the system within the hospital setting, considering privacy, and the positive feedback from patients and healthcare professionals, underscores the system's viability in a real-world clinical environment. The system employed in this study integrated physiological data from the EmbracePlus wristband and video footage from CCTV cameras, allowing for a comprehensive and Multimodal approach to AA detection. The EmbracePlus wristband system demonstrates promising results in detecting and classifying AA and pre-agitation events in individuals with severe dementia. The AA detection results are reflected in the video detection system, and the pre-agitation labels can be added to the system from EmbracePlus. The following discussion highlights key findings and their implications, followed by suggestions for future work.

The EmbracePlus wristband, leveraging both raw data and digital biomarkers, demonstrates its efficacy in discerning patterns associated with AA and pre-agitation. The personalized Extra Trees model emerged as the top-performing algorithm for the raw data, achieving high performance. We believe that the Extra Trees model randomness introduced during the creation of the trees often results in a more robust and accurate model. Extra Trees evaluates different features at each split, leading to a diverse set of decision trees. This diversity often enhances the model's ability to capture complex patterns in the data. Furthermore, features such as EDA tonic mean, accelerometer activity class, and pulse rate highlighted the significance of identifying AA patterns from raw data and digital biomarkers. Each participant had a different set of top features in AA identification. For participant \#1, the EDA Tonic mean showed high coloration to identify AA values, which indicates a connection between the emotional state of the patient and the observed AA. This supports the hypothesis that EDA, as a measure of sympathetic nervous system activity, is sensitive to emotional arousal. The top features for the second and third participants were the ones related to temperature and acceleration. These features can also have a strong correlation with AA since excessive movements and anger can contribute to the noted AA event. We also found pre-agitation label patterns for all participants from different features, at least six minutes before the actual AA event. The identification of pre-agitation patterns in the data suggests that physiological changes precede observable AA behavior. Being the first to explore these patterns in individuals with severe dementia from EmbracePlus wristband, this study lays the groundwork for a deeper understanding of the dynamics and physiological signatures of AA behaviors. The newfound ability to identify pre-agitation patterns offers a potential window for early intervention and preventive measures. 

Moreover, the video detection system, incorporating CCTV footage and advanced pose estimation techniques, was used along with the EmbracePlus wristband for AA detection. The privacy preservation technique, which follows the REB protocols in the hospital, does not exploit the patient’s personal features or body image without affecting the performance of the proposed model. The LSTM neural network and the GRU networks exhibited robust performances, achieving an accuracy rate of 98\% and 99\% respectively. We particularly favor the LSTM network for its faster inference time and smoother loss curve, which suggests a higher real-time performance. Moreover, the fast convergence of the LSTM model is likely because of its efficacy in learning longer sequences of AA present in our dataset. TThe LSTM model's high AUC of 99\% is particularly crucial in the context of healthcare, minimizing the risk of false negatives and ensuring that true AA events are accurately identified. Moreover, the fast convergence of the LSTM model is likely because of its efficacy in learning longer sequences of AA present in our dataset.  During the real-time deployment stage, the model adapts and continuously improves based on the collected data. The labeled pre-agitation labels collected from the wristband can be fed into the training set of the video detection system to provide insight into detecting pre-agitation from the video footage. The outcomes of our analysis are promising, demonstrating the potential of both LSTM and GRU neural networks in detecting AA in dementia care settings in real-time. However, expanding the dataset is necessary to solidify the current results.

For future work, we will focus on expanding our dataset by recruiting more participants from the hospital. We aim to validate the system over the long term, assessing its stability, generalizability, and adaptability to healthcare and home care environments. This step is crucial for building a database with a substantial number of AA and pre-agitation events, which is essential for developing a high-performance detection system using machine learning. Once our detection system is established, we plan to automate the real-time system capable of predicting AA. This involves receiving data in real-time, classifying the data, and sending notifications to the healthcare providers if an AA event occurs. In addition, we plan to improve the real-time AA detection of the video system. We also plan to use the pre-agitation labels from the EmbracePlus wristband to help our video detection model predict AA before they happen. During the initial stages of this study, healthcare providers will review and confirm all collected AA events. Their feedback is crucial in refining and enhancing the model's performance and should aid in identifying any limitations or challenges. Once the model is reliable, it will automatically detect and predict AA with no human intervention.

\section{Conclusion}

This study represents a notable step forward in developing an AA and pre-agitation detection system for individuals with severe dementia, employing a comprehensive approach integrating intelligent psychological biomarkers sensing  and video detection systems. We conducted a pilot study recruiting three participants from the Ontario Shores Center for Mental Health Sciences Institute. We used the EmbracePlus wristband for continuous health monitoring and video footage from CCTV cameras for real-time observation of AA events. In the preliminary data analysis, the raw data of the EmbracePlus wristband demonstrated exceptional performance in detecting AA events, with the Extra Trees model emerging as the top-performing algorithm for all the personalized models. For the general model, XGBoost outperform the rest of the models achieving an accuracy of 98\%. Exploring the digital biomarkers further strengthened the system's classification of AA, pre-agitation, and normal events. Pulse rate, activity class, and activity counts have emerged as critical indicators to detect AA. The study revealed the potential for detecting pre-agitation patterns, showcasing a six-minute lead time before actual AA events. This early detection capability holds promise for timely intervention and preventive measures. 

Aside from the EmbracePlus wristband, the video-based detection demonstrated promising results in detecting AA using LSTM, achieving a 98\% accuracy rate and a robust AUC of 99\%. The reported short inference time, of almost half of the GRU's inference time, and the fast learning are indicators of this model's capability to run in real-time. The high recall rate is particularly noteworthy, minimizing the risk of false negatives and accurately identifying actual AA events. This research presents a comprehensive methodology for detecting and predicting AA in individuals with severe dementia, combining wristband data and video detection. The preliminary data analysis's promising results underscore this Multimodal approach's potential to enhance patient care and safety by predicting AA events. This research will provide new directions for researchers interested in technologies for dementia care and provide challenging propositions in detecting and monitoring, modeling, and evaluating patient-specific interventions for PwD demonstrating NPS.

\bibliographystyle{IEEEtran}
\bibliography{references.bib}
\end{document}